\documentclass[conference,usenatbib]{basi}
\usepackage[T1]{fontenc}
\usepackage[british]{babel}
\usepackage[varg]{txfonts}
\usepackage{rotating}
\usepackage{dcolumn}

\begin{document}
\title[A long look at X-ray binaries]{Average spectral properties of galactic
X-ray binaries with 3 years of MAXI data}
\author[Islam et~al.]%
       {Nazma Islam$^{1,2}$\thanks{email: \texttt{nazma@rri.res.in}},
       Tatehiro Mihara$^3$, Mutsumi Sugizaki$^3$, Biswajit Paul$^1$ and Biman B. Nath$^1$ \\
       $^1$ Raman Research Institute,Sadashivanagar, Bangalore-560080, India\\
       $^2$ Joint Astronomy Programme, Indian Institute of Science, Bangalore-560012, India\\
       $^3$ MAXI team, RIKEN, 2-1 Hirosawa, Wako, Saitama 351-0198, Japan}

\pubyear{2013}
\volume{**}
\pagerange{**--**}

\date{Received --- ; accepted ---}

\maketitle

\label{firstpage}

\begin{abstract}
The energy spectra of X-ray binaries (XRBs) have been investigated during the last
few decades with many observatories in different energy bands and with
different energy resolutions.  However, these studies are carried out in
selected states of XRBs like during outbursts, transitions, quiescent
states, and are always done in limited time windows of pointed
observations. It is now possible to investigate the long term averaged spectra of a large number of 
X-ray binaries with the all sky monitor MAXI, which also has a broad energy band. 
Here we present the average spectral behaviour of a few representative XRBs. The long term averaged 
spectrum of Cyg X-1 is described by a sum of two power-laws having $\Gamma_{1} \sim$2.8 and $\Gamma_{2} \sim$1.2, 
along with a multi color disk blackbody having an inner disk temperature of 0.5 keV, GX 301-2 is described by a power-law 
with a high energy cut-off at $E_{c} \sim$ 15 keV and a blackbody component at 0.2 keV and that of Aql X-1 is 
described by a multi color disk blackbody at 2 keV and a power-law of $\Gamma \sim$ 2.2
We have also constructed the combined X-ray spectrum of the X-ray binaries in the Milky Way, which can 
be compared to the XRBs spectra of other galaxies observed with Chandra and XMM-Newton. 
These measurements are also relevant to investigate the X-ray interaction with the ISM and its 
contribution to the ionising X-ray background in the early universe.
\end{abstract}


\section{Introduction}

X-ray binaries have been the main object of study of many X-ray observatories. X-ray binaries have a wide range of spectral and 
temporal characteristics. While the long term temporal behaviour of some binaries are studied using all sky monitors 
like RXTE-ASM (for example, searching for periodicities \citep{wen}), 
the long term average spectral behaviour of 
these binaries have not been studied before. The reason is because these binaries have usually been observed only in their 
selected states like during outbursts, transitions and quiescent states \emph{i.e} in a limited time window of pointed observations. 
The wide field monitoring of the all sky with the Monitor of All Sky X-ray Image (MAXI), provides us the opportunity to investigate the 
long term averaged spectral behaviour of XRBs. \\
The main highlight of this work is the construction of long term averaged spectra of the bright X-ray binaries using 3 years (Msec) exposure of MAXI 
instrument. We have also constructed the combined X-ray spectrum of the X-ray binaries scaling by their distances to the 
galactic centre. This can be used in future studies on the X-ray interaction with the ISM and the contribution to the ionising X-ray 
background to the Epoch of Reionisation (EoR) in the early universe.
\section{Data Analysis}
MAXI is an all sky monitor operated on the International Space Station (ISS)\citep{matsuoka}. It has the best sensitivity and highest energy resolution 
amongst all operating all sky monitors. It scans the entire sky every 92 min as ISS follows its orbit. 
It consists of two X-ray cameras: Gas Slit Camera (GSC) \citep{mihara,sugizaki}, operating between 2-20 keV and Solid state Slit Camera (SSC) 
\citep{tomida,tsunemi}, 
operating between 0.7 to 7 keV. GSC consists of six units of large area position sensitive Xenon proportional counters with slit and slat collimators, 
to cover 1.5$^\circ \times$ 160$^\circ$ FOV. The typical daily exposure of GSC is 1500 cm$^{2}$ s for one source.  
We have used MAXI on demand process\footnote {http://maxi.riken.jp/mxondem} to analyze long term data of sources. 
The long term averaged spectra were extracted from MJD:55058 to MJD:56206, 
for all the 65 XRBs included in our sample and were fitted using xspec \emph{v:12.6}. Long term averaged spectra for a few of the sources 
are described in the following subsections.
\subsection{Cyg X-1}
The galactic black hole high mass X-ray binary Cyg X-1 has two distinct spectral states: high soft and low hard states \citep{remillard}. 
High soft state is dominated by thermal X-ray spectrum from the accretion disk whereas the low hard state consists of a power-law spectrum. 
The typical photon index $\Gamma$ of the hard component in the high soft state
 and that in the low hard state is in the range of 1.7$\sim$2.5. 
Cyg X-1 was continuously in a low hard state from MJD:55058 to MJD:55300 and was mostly in high soft state from MJD:55800 to MJD:56206 
(except for a short hard state from MJD:55912 to MJD:55941) as was observed in the light curves with SWIFT BAT, MAXI and partly with RXTE ASM.
The averaged low hard state spectra extracted from MJD:55058 to MJD:55300 is described by a power-law with a photon index $\Gamma_{1} \sim$1.6. 
The averaged high soft state spectra extracted from MJD:55800 to MJD:56206 is 
described as a sum of a multicolor disk blackbody, having an inner disk temperature of 0.6 keV, and a power-law with a photon index $\Gamma_{1} \sim$2.4.
The long term averaged spectrum of Cyg X-1 over both the above spectral states is described as sum of two power-laws and a 
multi color disk blackbody model (Fig.~\ref{graph}). The two power-laws have photon indices of $\Gamma_{1} \sim$2.8 and $\Gamma_{2} \sim$1.2. 
The disk blackbody has an inner disk temperature of 0.5 keV. A broad Fe emission line at 6.6 keV is also present in the spectra. 
It has an equivalent width of $\sim$ 0.32 keV.
\subsection{GX 301-2}
GX 301-2 is an wind fed accreting high mass X-ray pulsar having a spin period of $\sim$ 685 sec and an orbital period of 41.5 days around a B type supergiant 
companion. It is a strongly variable source and has a large circumstellar absorption column density around the neutron star. 
The X-ray spectrum from an accretion mound and hot spots have been modelled with a power-law with high energy cut-off \citep{nagase}.
The long term averaged X-ray spectrum of GX 301-2 is described by a power-law having $\Gamma \sim$ 0.3 with a high energy cut-off at 
$E_{c} \sim$ 15 keV, with the presence of an additional blackbody component at 0.2 keV (Fig.~\ref{graph}). 
A single absorber model (phabs) is used which has 
$N_{H} \sim 10^{23}$ atoms cm$^{-2}$. A strong Fe emission line at 6.4 keV is also present in the spectra. 
It has an equivalent width of $\sim$ 0.74 keV.
\subsection{Aql X-1}
\begin{figure}
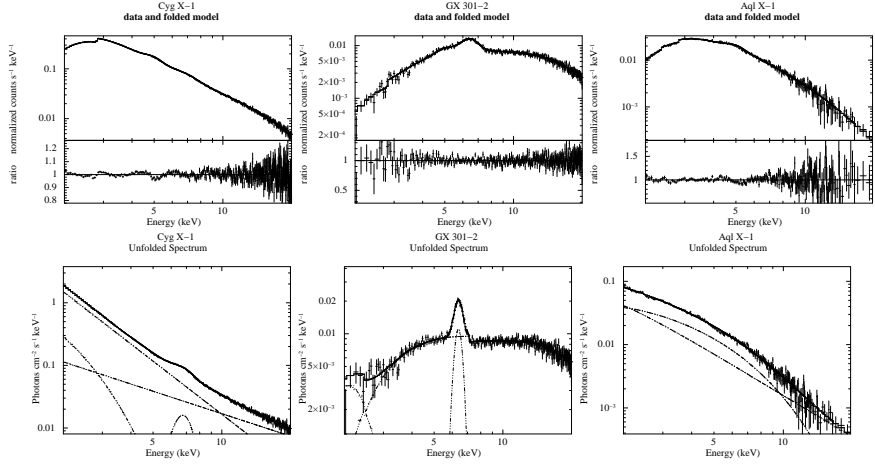

\centering
\includegraphics[angle=-90,scale=0.14]{islam_01.ps}
\includegraphics[angle=-90,scale=0.14]{islam_02.ps}
\includegraphics[angle=-90,scale=0.14]{islam_03.ps}
\includegraphics[angle=-90,scale=0.14]{islam_04.ps}
\includegraphics[angle=-90,scale=0.14]{islam_05.ps}
\includegraphics[angle=-90,scale=0.14]{islam_06.ps}
\caption{The observed X-ray spectrum along with the best fit model, ratio between the data and the model and the unfolded spectrum 
shown for Cyg X-1 (left panel), GX 301-2 (middle panel) and Aql X-1 (right panel).}
\label{graph}
\end{figure}

Aql X-1 is a low mass X-ray binary transient, where the mass transfer occurs through Roche lobe overflow. 
Within 3 years of MAXI operation, it has undergone 3 outbursts \citep{asai}. It is classified as an atoll source having three distinct spectral states: 
High soft (banana state), Intermediate (Island state), and Low hard (Extreme Island state), on the basis of the tracks traced by the source 
on the color-color diagrams. The soft spectral states of LMXBs has been modelled with a multi color disk black body and hard spectral state have 
been modelled with a power-law having $\Gamma \sim$ 2 \citep{barret}. 
The long term averaged spectrum of Aql X-1 is described by a multi color disk blackbody of 2 keV and a power-law of $\Gamma \sim$ 2.2 (Fig.~\ref{graph}).

\begin{figure}
\centering
\includegraphics[angle=-90,scale=0.22]{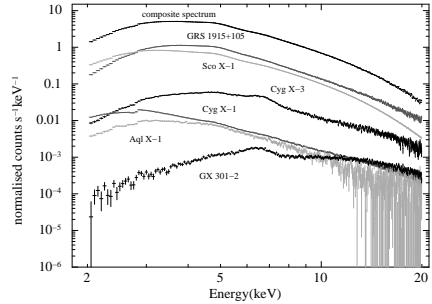}
\caption{Composite spectrum of X-ray binaries in Milky Way, compared with distance scaled spectra of few bright binaries like GRS 1915+105, Sco X-1, 
Cyg X-1, Cyg X-3, Aql X-1 and GX 301-2.}
\label{composite}
\end{figure}

\subsection{Composite X ray spectrum of Milky Way}
To construct the composite spectrum of X-ray binaries, these long term averaged spectra of 65 X-ray binaries are 
scaled for a distance of 8.5 kpc, same as that of the galactic centre. 
These scaled spectra are added to give a composite X-ray spectrum of Milky Way for an observer situated outside the Galaxy. 
For the transient sources like Aql X-1, we have used all the time-averaged data including both outbursts and quiescence.
As seen in Fig.(~\ref{composite}), the spectra of GRS 1915+105 and Sco X-1, scaled to their distances, contributes to a 
significant fraction of the X ray binaries composite spectrum. The total luminosity of the composite X ray spectrum of Milky Way 
L$_{X} \sim 2.5\times10^{39}$ erg s$^{-1}$.

\section{Conclusions and future implications}
We have constructed long term averaged spectra of 65 X-ray binaries, using 3 years of MAXI data. 
The averaged spectral behaviour of these binaries are related to the binary type. For example, all the accretion powered pulsars are well described by a 
power-law with a highecut model. These in turn, shed light on their emission mechanisms and accretion geometry. \\
The composite X-ray binaries spectrum of Milky Way has a direct relation to the X-rays interaction with ISM, which will be investigated in future 
studies. The X-rays from primordial X ray binaries (from first stars) contribute significantly to the ionising X-ray background at high redshift. 
The hard X-rays have an important contribution to the re-ionisation of cosmic abundance of neutral hydrogen since they have longer mean free path 
than UV photons \citep{venkatesan}. Many authors \citep{mirabel,power2013} have tried to estimate the contribution of X-rays in 
epoch of re-ionisation (EoR), by assuming a template spectra that is used in re-ionisation simulations. 
With the X-ray binaries composite spectra of Milky Way Fig.(~\ref{composite}), it is possible to 
have realistic estimates on the fraction contributional of hard X-rays in heating of IGM and EoR.
\section{Acknowledgements}
This research has made use of MAXI data provided by RIKEN, JAXA and the MAXI team. The authors would like to thank S. Nakahira for maintaining 
the MAXI on demand process. We also thank Prof. Matsuoka for his useful comments. NI thank RIKEN, where this project was initiated.

\end{document}